%% file: 00-main.tex
\documentclass[conference,a4paper]{IEEEtran}
\IEEEoverridecommandlockouts
\usepackage{url}
\usepackage{multicol}
\usepackage{multirow}
\usepackage{amssymb}
\usepackage{pifont}
\usepackage{amsmath,amssymb,amsfonts}
\usepackage{amsmath,amssymb,amsfonts,color,soul}
\usepackage[ruled,vlined,linesnumbered]{algorithm2e}
\usepackage{algpseudocode}
\usepackage[flushleft]{threeparttable}
\usepackage{graphicx}
\usepackage{textcomp}
\usepackage{xcolor}
\DeclareMathOperator{\E}{\mathbb{E}}
\def\BibTeX{{\rm B\kern-.05em{\sc i\kern-.025em b}\kern-.08em
    T\kern-.1667em\lower.7ex\hbox{E}\kern-.125emX}}

\usepackage[nolist]{acronym}
\acused{wifi}
\acused{wlan}
\acused{gnss}

\usepackage[backend=biber,style=ieee,  sortcites=true,isbn=false, doi=false, url=false,mincitenames=1, maxcitenames=1,maxbibnames=3,hyperref=false]{biblatex}
\usepackage{balance}

\usepackage{soul}
\soulregister\cite7
\soulregister\ref7
\soulregister\SI7
\soulregister\ac7

\usepackage[detect-all=true, binary-units=true]{siunitx}
\DeclareSIUnit\decibelm{dBm}
    
\usepackage[detect-all=true, binary-units=true]{siunitx}
\usepackage{array}
\usepackage{booktabs}
\usepackage{mathtools}

\DeclarePairedDelimiter\abs{\lvert}{\rvert}%

\newcommand{\MYheader}{2022 International Conference on Indoor Positioning and Indoor Navigation (IPIN), 5 – 7 Sep. 2022, Beijing, China}
\newcommand{\MYcopyrigth}{978-1-7281-6218-8/22/\$31.00~\copyright~2022~IEEE\hfill}

\renewcommand{\MYheader}{}
\renewcommand{\MYcopyrigth}{}

\makeatletter
\def\ps@headings{%
	\def\@oddhead{\hfill\MYheader\hfill }
	\def\@evenhead{\hfill\MYheader\hfill}
	\def\@oddfoot{}%
	\def\@evenfoot{}}
\def\ps@IEEEtitlepagestyle{%
	\def\@oddhead{\hfill\MYheader\hfill}
	\def\@evenhead{\hfill\MYheader\hfill}
	\def\@oddfoot{\MYcopyrigth}%
	\def\@evenfoot{\MYcopyrigth}
    }
\makeatother
\begin{document}

\title{SURIMI: Supervised Radio Map Augmentation with Deep Learning and a Generative Adversarial Network for Fingerprint-based Indoor Positioning

\thanks{%
Corresponding Author: D. Quezada Gaibor (\texttt{quezada@uji.com})
}

\thanks{The authors gratefully acknowledge funding from European Union’s Horizon 2020 Research and Innovation programme under the Marie Sk\l{}odowska Curie grant agreements No.~$813278$ (A-WEAR: A network for dynamic wearable applications with privacy constraints, {http://www.a-wear.eu/}) and No.~$101023072$ (ORIENTATE: Low-cost Reliable Indoor Positioning in Smart Factories, {http://orientate.dsi.uminho.pt}).}
}

\author{%
\IEEEauthorblockN{%
Darwin Quezada-Gaibor%
\IEEEauthorrefmark{1}\textsuperscript{,}\IEEEauthorrefmark{2}, %
Joaquín Torres-Sospedra%
\IEEEauthorrefmark{3}, %
\\Jari Nurmi%
\IEEEauthorrefmark{2}, %
Yevgeni Koucheryavy%
\IEEEauthorrefmark{2}, %
and Joaquín Huerta%
\IEEEauthorrefmark{1}%
}

\IEEEauthorblockA{\IEEEauthorrefmark{1}\textit{Institute of New Imaging Technologies}, \textit{Universitat Jaume I}, Castellón, Spain}
\IEEEauthorblockA{\IEEEauthorrefmark{2}\textit{Electrical Engineering Unit}, \textit{Tampere University}, Tampere, Finland}
\IEEEauthorblockA{\IEEEauthorrefmark{3}\textit{Algoritmi Research Centre}, \textit{Universidade do Minho}, Guimarães, Portugal}
}

\maketitle

\begin{abstract}
Indoor Positioning based on Machine Learning has drawn increasing attention both in the academy and the industry as meaningful information from the reference data can be extracted.
Many researchers are using supervised, semi-supervised, and unsupervised Machine Learning models to reduce the positioning error and offer reliable solutions to the end-users. In this article, we propose a new architecture by combining \ac{cnn}, \ac{lstm} and \ac{gan} in order to increase the training data and thus improve the position accuracy. The proposed combination of supervised and unsupervised models was tested in 17 public datasets, providing an extensive analysis of its performance. As a result, the positioning error has been reduced in more than $70\%$ of them.
\end{abstract}


\input{acronyms}

\input{01-introduction}
\input{02-relatedwork}
\input{03-proposed-method}

\input{04-experiments-results}

\input{05-discussion}
\input{06-conclusions}


\renewcommand*{\UrlFont}{\rmfamily}\printbibliography

\balance

\end{document}

%% file: acronyms.tex
\begin{acronym}[XXX] 
\acro{ap}[AP]{Access Point}
\acro{apc}[APC]{affinity propagation clustering}
\acro{ble}[BLE]{Bluetooth Low Energy}
\acro{cnn}[CNN]{Convolutional Neural Network}
\acro{csi}[CSI]{Channel State Information}
\acro{dbscan}[DBSCAN]{Density-based Spatial Clustering of Applications with Noise}
\acro{fp}[FP]{fingerprinting}
\acro{fpc}[FPC]{fingerprinting clustering}
\acro{gan}[GAN]{Generative Adversarial Network}
\acro{iot}[IoT]{Internet of Things}
\acro{ips}[IPS]{Indoor Positioning System}
\acro{knn}[$k$-NN]{k-nearest neighbors}
\acro{leakyrelu}[LeakyReLU]{Leaky Rectified Linear Unit}
\acro{lda}[LDA]{Linear Discriminant Analysis}
\acro{lbs}[LBS]{location-based service}
\acro{lstm}[LSTM]{Long short-term memory}
\acro{mac}[MAC]{Media Access Control}
\acro{ml}[ML]{Machine Learning}
\acro{mlp}[MLP]{multilayer perceptron}
\acro{nn}[NN]{Nearest Neighbour}
\acro{pca}[PCA]{Principal Component Analysis}
\acro{relu}[ReLU]{rectified linear}
\acro{rf}[RF]{Radio Frequency}
\acro{rnn}[RNN]{recurrent neural networks}
\acro{rp}[RP]{Reference Point}
\acro{rs}[RS]{Recommender Systems}
\acro{rss}[RSS]{Received Signal Strength}
\acro{sae}[SAE]{Stacked Auto-Encoder}
\acro{tsne}[t-SNE]{T-distributed Stochastic Neighbor Embedding}
\acro{uwb}[UWB]{ultra-wideband}
\acro{vlc}[VLC]{Visible light communication}
\acro{wifi}[\mbox{Wi-Fi}]{IEEE 802.11 Wireless LAN}
\acro{wknn}[WKNN]{weighted k-nearest
neighbor}
\acro{wlan}[WLAN]{Wireless LAN}
\acro{wsn}[WSN]{Wireless Sensors Networks}

\end{acronym}

%% file: 01-introduction.tex
\section{Introduction}
\label{sec:introduction}

Since the early years of the 21st century special emphasis is placed on the design of \acp{ips} based on \acf{ml} models and \ac{wifi} fingerprinting \cite{torres2020new,yan2021elm,lu2018indoor} which include supervised, semi-supervised, unsupervised and reinforcement learning models \cite{asmar2004smartslam, miura2006adequate}. 
Although \ac{wifi} fingerprinting is very popular, it suffers from scalability problems in terms of position accuracy and time response \cite{mao2018scalability}.

In general, \ac{wifi} Fingerprinting is divided into two well-defined phases: the offline phase where the radio map is generated by collecting fingerprints at known reference points, and the online phase where the incoming fingerprint (at an unknown position) is compared to the fingerprints in the radio map in order to estimate the device position \cite{dai2019autonomous}. Therefore, in the traditional fingerprinting approach~\cite{radar}, the computational cost of the latter one is highly dependent on the radio map size.

Some researchers have proposed alternatives balancing between the positioning accuracy and execution time and, thus, diminishing the scalability problems in \ac{wifi} fingerprinting. For instance, \textcite{song2019cnnloc} proposed a new solution for \ac{wifi} fingerprinting based on the combination of \ac{sae} and \ac{cnn}, to improve the accuracy of the building and floor detection, and therefore, the position accuracy. \textcite{torres2020comprehensive} provided an analysis of clustering algorithms in \ac{wifi} fingerprinting --in terms of positioning accuracy and computational costs--  which included an evaluation framework with many open-source radio maps. 

It is well-known that creating an accurate radio-map is time-consuming and requires extensive manual-labour. However, the radio map should be updated after any significant change in the environment to guarantee the position accuracy \cite{borras2016indoor}. An alternative to reduce the time for a radio map collection is to generate \emph{artificial} fingerprints using mathematical or statistical models \cite{alshami2017adaptive}.
Currently, some \ac{ml} models (e.g., \acp{gan}) are capable of generating new data, including artificial \ac{rss} values for radio map augmentation~\cite{njima2021indoor}, which cannot be easy to identify whether they are real or not. 

In this article we propose the combination of deep learning models (\ac{cnn} and \ac{lstm}) and conditional \ac{gan}, in order to enhance the radio map without additional manual labour, and therefore, to reduce the positioning and localization error. The main contributions of this work are the following:

\begin{itemize}
  \item A new generalized machine learning model for data augmentation and indoor positioning (\ac{cnn}-\ac{lstm}-c\ac{gan}).
  \item An extended analysis of the proposed model with several \ac{wifi} and \ac{ble} datasets.
  \item The source-code of the \ac{cnn}-\ac{lstm}-c\ac{gan} model \cite{quezada2021surimi}.
\end{itemize}

The remaining of this paper is organized as follows.
Section \ref{sec:related_work} provides a general overview of indoor positioning based on deep learning models. 
Section \ref{sec:model} describes the proposed deep learning model for \ac{wifi} fingerprinting. 
Section \ref{sec:usecases} provides the details of experiments carried out and their results. Section~\ref{sec:discussion} offers a general discussion of the results. Finally,
section \ref{sec:conclusions} provides the main conclusions of this research article.

%% file: 02-relatedwork.tex
\section{Related work}
\label{sec:related_work}

\ac{wifi} fingerprinting technique has been widely used for indoor positioning purpose due to its low-cost and also the support of modern user devices such as smartphones, wearable and \ac{iot} devices to \ac{wifi} technology \cite{radhakrishnan2015smartphones,dinh2020smartphone}. However, \ac{wifi} fingerprinting is not as accurate as \ac{uwb} or \ac{vlc}, which can reach centimetre level accuracy \cite{li2018vlc,flueratoru2021highaccuracy}. With the aim of reducing the positioning error and enhance the scalability of \ac{wifi} fingerprinting some authors have proposed multiple solutions that contemplate the use of deep learning models such as \ac{rnn}, \ac{cnn}, \ac{lstm}, among others \cite{bai2020dl,chen2020novel}.

Deep learning models are specially used for feature learning, allowing the extraction of meaningful information from the raw data. In this case, \ac{cnn}, \ac{lstm} or other models are used to extract the main information of the radio map and then estimate the user's localization. For instance, \textcite{song2019novel} combined \ac{sae} and \ac{cnn} to provide a highly accurate indoor localisation solution which achieved $100\%$ of the building rate and $95\%$ of the floor rate approximately. The authors used the \ac{sae} to extract key characteristics of the \ac{rss} values in the dataset, and the \ac{cnn} model for high accurate classification.

Recent models such as \acp{gan} \cite{mirza2014conditional} are not only used for generating new images or text but also, these models are used to generate new \ac{rss} values on the basis of real data. \textcite{belmonte2020recurrent} offered a novel framework which incorporates the advantages of the \ac{rnn} and \ac{gan} in order to recover the information when the measurements are not received from the devices deployed in the environment. The recovered information is used to compute the user path and update the radio map. As a result, the positioning error was reduced by $3\%$ in comparison to other techniques.

\textcite{li2021af} proposed a new model namely Amplitude-Feature Deep Convolutional \ac{gan} (AF-DCGAN). In this research work, the authors used the \ac{csi} to enhance the radio map, and transformed each amplitude feature map into an image for further processing. Thereby, the proposed model was used to generate new images based on the original image and these new amplitude feature maps are added to the original one to reduce the positioning error.

\textcite{njima2021indoor} used deep learning models and \ac{gan} to generate new \ac{rss} data in order to extend the radio map. Unlike the earlier research, work published in \cite{njima2021indoor} has established a new criteria to select the most realistic fake fingerprints also known as artificial or synthetic fingerprints. As a result, the authors have shown  an improvement in localization accuracy of $15\%$ in comparison with the benchmark.

This paper offers a new framework that includes the positioning and \ac{gan} models to generate realistic fingerprints and improve the training stage in the offline phase of \ac{wifi} fingerprinting. Unlike the previous research work, the \ac{gan} model consists of a conditional label (floor label) used as an extra parameter in the generative model, allowing to generate artificial data for multi-building and multi-floor environments.

%% file: 03-proposed-method.tex
\section{Data Augmentation Framework}
\label{sec:model}
This section provides a general description of the proposed framework and their components.  

\subsection{Indoor Positioning Framework description}
\begin{figure*}[!hbt]
    \centering
    \includegraphics[width=\textwidth]{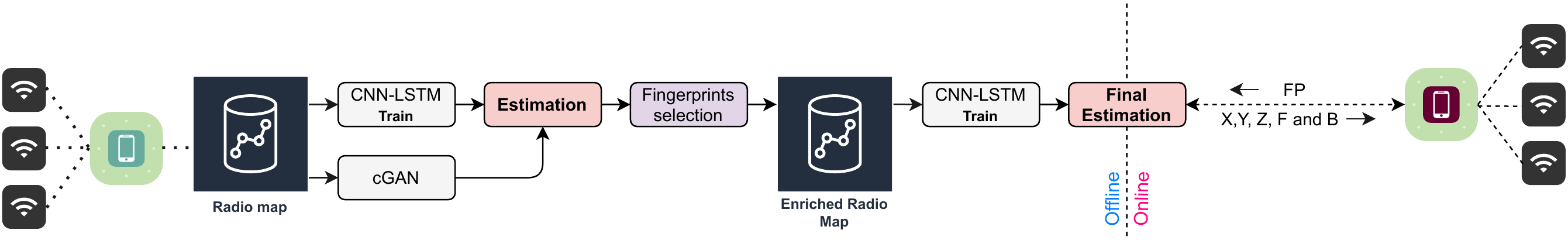}
    \caption{\ac{cnn}-\ac{lstm}-c\ac{gan} \ac{ble}/\ac{wifi} fingerprinting process}
    \label{fig:fingerprinting}
\end{figure*}

The proposed indoor positioning framework based on \ac{wifi} fingerprinting is devoted to increase the fingerprints in the radio map and reduce the positioning error using three well-known machine learning models \ac{cnn}, \ac{lstm} and c\ac{gan}. The \ac{cnn}-\ac{lstm} model is used to predict the 2D/3D position, including floor and building, while the c\ac{gan} is used to generate \emph{artificial} fingerprints to enrich the radio map. These \ac{ml} models use only the information stored in the radio map, which include the \ac{rss} values, X, Y and Z position, and floor \& building tags. Once the artificial fingerprints are generated the position for each fingerprint is predicted and only the most relevant fingerprints are added to the original dataset to form an enriched radio map (see Fig.~\ref{fig:fingerprinting}).

\subsection{Fingerprinting}
Indoor positioning based on \ac{wifi} fingerprinting consists of two main phases as mentioned in earlier paragraphs. In the offline stage, the radio map is formed and the proposed machine learning model is trained in order to be used in the next stage. In the online phase, the user position is predicted.

Generally, a radio map ($\Psi$) is formed by $m$ number of samples -- fingerprints -- and $n$ number of \acp{ap} ($m \times n$). Each position in the radio map is corresponding to a \ac{rss} value ($\psi_{ij}$) in the \textit{i}-th position or sample ($ i=1,2,...,m$) and transmitted by the n-AP ($j=1,2,...,n$).
$$
\Psi=
\begin{bmatrix}
\psi_{11}, \psi_{12} & \hdots & \psi_{1n}\\
\psi_{21}, \psi_{22}  & \hdots & \psi_{2n}\\
\vdots & \ddots & \vdots\\
\psi_{m1} & \hdots & \psi_{mn}
\end{bmatrix}
$$
Each sample (fingerprint) in the dataset is linked to a known training position (X, Y, Z, floor and building), these labels are used during the training stage. 

\subsection{Data preprocessing}

Data preproccesing is fundamental in the early stages of data analysis and machine learning, using multiple techniques for data cleaning, scaling, encoding, among other. Therefore, before data processing we have changed the data representation of each dataset as is suggested in \cite{torres2015comprehensive,torres2020comprehensive}. The new data representation reduce the data complexity allowing to extract more useful information with \ac{ml} models. Additionally, \emph{MinMaxScaler} is used to scale the longitude, latitude and altitude labels -- X, Y and Z -- (see Eq.~\ref{eq:minmax}) and \emph{OneHotEncoder} for floor and building.
\begin{equation}
\label{eq:minmax}
    X_{scaler} = \frac{X-X_{min}}{X_{max} - X_{min}}
\end{equation}
where, $X_{min}$ represents the minimum value and $X_{max}$ is the maximum value in the labels of longitude, latitude and altitude. 

\subsection{Positioning model - CNN-LSTM} 

Given the heterogeneity of indoor environments, user devices and the fluctuation of the \ac{wifi} signal over the time makes deep learning models useful to learn those patterns or changes \cite{njima2021indoor}. The first layer of the proposed \ac{cnn}-\ac{lstm} model is composed of a \emph{Conv1D} layer that helps to extract spatial characteristics of the input data. Then, we have a \emph{max pooling} layer (MaxPooling1D), which allows computing the maximum value for each patch in the feature map. To reduce overfitting problems, a \emph{dropout} layer is added to the positioning model, the dropout rate is $0.5$ for all the models. The dropout layer is used to deactivate some neurons during the training stage. The next three layers are a Conv1D, MaxPooling1D and dropout with similar characteristics to the previous layers. The following layer is the \emph{flatten} layer that converts the data into a 1-dimensional array. The next layer is the \emph{\ac{lstm}} used to learn order dependence. Finally, we have the \emph{dense} layer fully connected which perform the prediction of the 2D/3D position (longitude, latitude and altitude).

The model to classify the fingerprints per floor is similar to the positioning model, but with different values for the filter, activation function, and learning rate, among others. In the case of the building model, it is less complex in comparison with the previous models and it uses only one convolutional layer.
%
Table \ref{table:parameters} shows the positioning model layers with their parameters and values. Additionally, we adopt early stopping ($patience=5$) to avoid overfitting.

\input{tables/tbl_parameters}

\subsection{Data augmentation}

\begin{figure}[!hbt]
    \centering
    \includegraphics[width=\columnwidth]{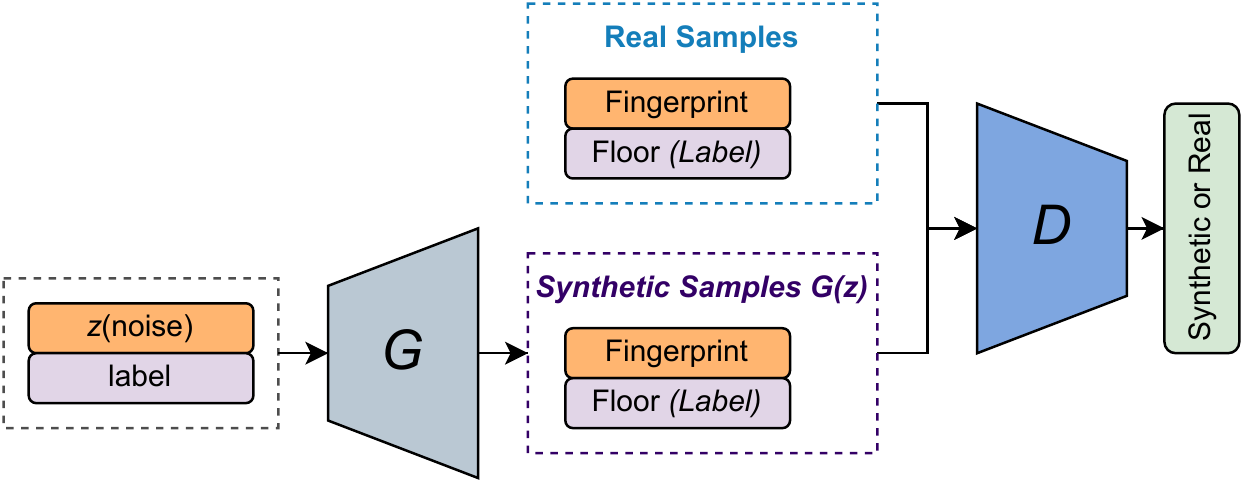}
    \caption{c\ac{gan} architecture}
    \label{fig:cganmodel}
\end{figure}

Data augmentation is done by using the c\ac{gan} architecture, which contains two main components (models), the discriminator and generator. Unlike \ac{gan}, c\ac{gan} includes a conditional label (class) allowing the targeted generation of fingerprints of a given floor or building. The primary function of the discriminator is to distinguish or classify the fingerprints as either artificial or real (i.e., it is a binary classifier of either $0$ or $1$). The generator is devoted to generating new fingerprints similar to real fingerprints (see Fig.~\ref{fig:cganmodel}).

In this architecture, both models (discriminator and generator) are trained together in an adversarial manner. It means that the discriminator capabilities increase at the expense of decreasing the capabilities of the generator and vice-versa also known as min-max game (see Eq.~\ref{eq:cgan}). As a result, new original fingerprints are generated in locations close to real ones.

\begin{equation}
    \label{eq:cgan}
    \begin{aligned}
    \min_{G} \max_{D}V(D,G) = &  \E_{\Psi \sim p_{data}(\Psi)}[\log{D(\Psi|y)}] +\\
    & \E_{z \sim p_{z}(z)}[\log{1-D(G(\Psi|y))}]
    \end{aligned}
\end{equation}
where, $V(D,G)$ is the value function, D represents the discriminator and G is the generator. \textbf{$x$} represents the fingerprints, \textbf{$y$} the classes, \textbf{$z$} the noise values. $p_{data}(x)$ is the data distribution over the data $x$ and $p_{z}(z)$ is the noise distribution.

\begin{figure}[!hbt]
    \centering
    \includegraphics[width=\columnwidth]{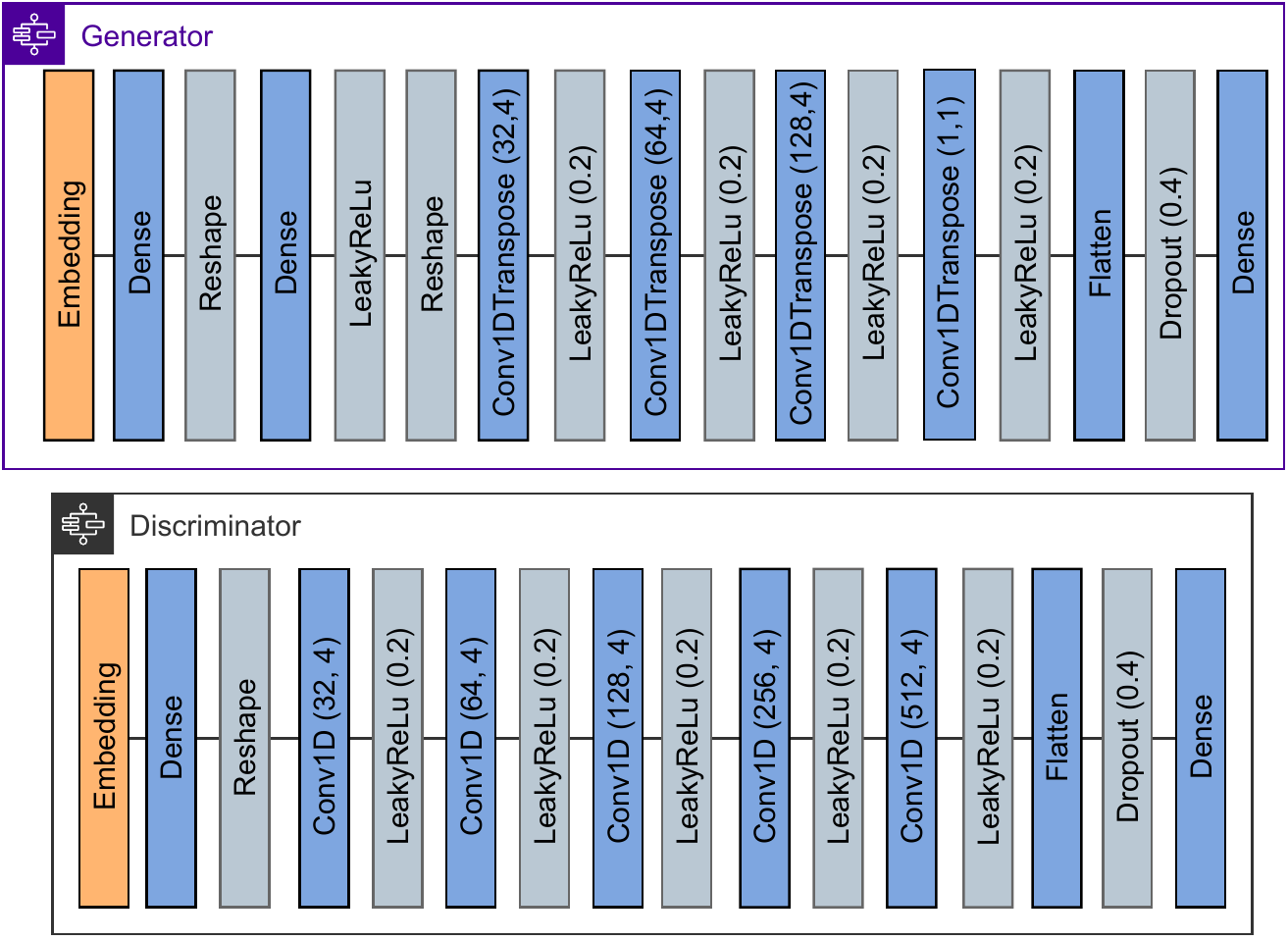}
    \caption{Discriminator and generator model}
    \label{fig:discgen}
\end{figure}

Fig. \ref{fig:discgen} shows the layers of the discriminator and generator model together with some of their parameters. These \ac{ml} models contain convolutional layers (Conv1D), dense layers and flatten layers. In the same way as the positioning model, dropout layers were added to each model in order to reduce the overfitting. Unlike the discriminator, the generator model uses Conv1DTranspose instead of Conv1D, which is the inverse operation of Conv1D.  Additionally, both the discriminator and generator use \ac{leakyrelu} activation function. LeakyReLu is mathematically defined as (see Eq.~\ref{eq:leakyrelu}) \cite{xu2015empirical}.
\begin{equation}
    \label{eq:leakyrelu}
    \begin{aligned}
     \mu_{i} & =\begin{cases} 
        x_i, & \text{if } x_i \geq 0, \\
        \frac{x_i}{a_i}, & \text{if } x_i < 0
    \end{cases}
    \end{aligned}
\end{equation}

where, $\mu_{i}$ represents the output after passing the LeakyReLu function, $a_i$ is a fixed parameter between 1 to $\infty$ and $x_i$ denotes the input data.

The following parameters were used to compile both the discriminator and generator model: Adam optimizer, learning rate equal to 0.0002 and binary cross-entropy loss.

\subsection{Fingerprints selection}

Given that the position of the new fingerprints is unknown and can not be parameterized during the training stage, a new algorithm is designed to select the most relevant synthetic fingerprints. The c\ac{gan} network thus generates $n$ number of new fingerprints and then the proposed algorithm selects only those fingerprints close to the real ones (see Algorithm~\ref{alg:fp_class_balancing}). Once the fingerprints are generated and selected, the \ac{cnn}-\ac{lstm} model is trained with the augmented radio map to increase the prediction and classification accuracy.

The proposed algorithm requires five input parameters: the training fingerprints ($\Psi_{XTR}$) and their labels ($\Psi_{yTR}$), the number of artificial fingerprints to be generated ($\aleph_{nfs}$), and a list of distances ($\aleph_d$) which is used to select the most relevant fingerprints after $n$ number of iterations ($\aleph_i$). For each iteration, the algorithm generates a latent space (i.e., the latent space is a Gaussian distribution with zero mean and standard deviation of one). In this case, the latent space is a $\aleph_{nfs} \times n$ matrix ($\iota_p$) ($n$ is equal to the number of features in the radio map) and their corresponding labels ($\iota_l$). The latent space is used to generate the synthetic fingerprints ($\Psi'_{XF}$). Once the new fingerprints are generated, the \ac{cnn}-\ac{lstm} is used to predict their positions (X, Y, Z, floor and building) ($\Psi'_{yF}$). In order to select the most relevant fingerprints, the distance between the positions of artificial and real fingerprints ($\Psi'_{yF}$ and $\Psi'_{yTR}$ respectively) is computed in order to create a distance matrix ($D_{ij} \in \mathcal{R}^{\aleph_{nfs} \times m}$, where $m$ is the number of samples in the training dataset). This distance matrix help us to determine whether there is at least one artificial fingerprint close to a real one ($\Psi_{XFi} \leftarrow \Psi'_{XFi} \Leftrightarrow \exists d_{ij}, d_{ij}  \leqslant dist$, where $d_{ij}$ is the computed distance in the i-$th$ and j-$th$ position in the distance matrix and $dist$ is the distance used to select the artificial fingerprints)). Thus, only those fingerprints in the established range will be selected to enrich the radio map.

The output parameters are the artificial fingerprints ($\Psi_{XF}$) and their corresponding labels ($\Psi_{yF}$).


\SetKwComment{Comment}{/* }{ */}
\begin{algorithm}[!hbt]
\SetAlgoLined
\DontPrintSemicolon
\KwIn{$\Psi_{XTR}$, $\Psi_{yTR}$, $\aleph_d$, $\aleph_{nfs}$, $\aleph_i$, }
\KwOut{$\Psi_{FX}$, $\Psi_{Fy}$}
\SetKwFunction{FMain}{artificialFPSelection}
    \SetKwProg{Fn}{Function}{:}{}
    \Fn{\FMain{$\Psi_{XTR}$, $\Psi_{yTR}$, $\aleph_d$, $\aleph_i$, $\aleph_{nfs}$}}{
        \For{$dist$ in $\aleph_d$}{
            \For{$iter$ = 0 to $\aleph_i$}{
                \tcc{Latent space $\iota_p$}
            $\iota_p \in \mathcal{R}^{\aleph_{nfs} \times n}$\\
            
            \tcc{Labels}
            $\iota_l \in \mathcal{R}^{\aleph_{nfs}}$\\
            
            $\Psi'_{XF}$ = cgan.predict($[\iota_p, \iota_l]$)\\
            \tcc{Predict position (X,Y,Z), floor and building ($\Psi'_{yF}$)}
            $\Psi'_{yF}$ = cnn\_lstm.predict($\Psi'_{XF}$)\\
            \tcc{Distance matrix between the real and artificial fingerprints.}
            $D_{ij} = D(\Psi'_{yF}, \Psi'_{yTR}) = \Vert \Psi'_{yF} - \Psi'_{yTR} \Vert ^2$\\
            $D_{ij} \in \mathcal{R}^{\aleph_{nfs} \times m}$\\
            \tcc{Selecting relevant artificial fingerprints}
            $\Psi_{XFi} \leftarrow \Psi'_{XFi} \Leftrightarrow \exists d_{ij}, d_{ij}  \leqslant dist$
            }
        }
    \textbf{return} $\Psi_{XF}$, $\Psi_{yF}$
}
\textbf{End Function}
\caption{Fingerprints selection}
\label{alg:fp_class_balancing}
\end{algorithm}

\subsection{Training Method Description}
Three different methods have been used to train the c\ac{gan} network in order to identify the optimal parameters to generate new fingerprints. 

\subsubsection{First Method (M1)} This method consists of training the model per building using floor label as the conditional parameter in the c\ac{gan} network.
\subsubsection{Second Method (M2)} In this method, the training stage is done per floor and the conditional parameter is the building label. 
\subsubsection{Third Method (M3)} This method uses the whole dataset to train the c\ac{gan} network, and the conditional label is the floor. Unlike of the previous methods, this third method does not take into account the building label during the c\ac{gan} training stage.

%% file: tables/tbl_parameters.tex
\begin{table}[!hbtp]
    \tabcolsep 4pt
    \caption{CNN-LSTM Parameter}
    \label{table:parameters}
    \centering
    \begin{threeparttable}
    \begin{tabular}{
    l
    l
    l
    l
    l
    l
    }
         \toprule
            &Layer
            &Parameter
            &Position
            &Floor
            &Building\\
         \midrule
\parbox[t]{2mm}{\multirow{12}{*}{\rotatebox[origin=c]{90}{TimeDistributed}}} &\multirow{3}{*}{Conv1D}&Filter&8&16&16\\
&&Activation&elu&relu&relu\\
&&kernel\_size&1&1&1\\
\cmidrule{2-6}
&MaxPooling1D&pool\_size&1&2&2\\
\cmidrule{2-6}
&Dropout&rate&0.5&0.5&0.5\\
\cmidrule{2-6}
&\multirow{4}{*}{Conv1D}&Filter&8&32&\\
&&Activation&elu&relu&\\
&&kernel\_size&1&1&\\
&&padding&same&same&\\
\cmidrule{2-6}
&MaxPooling1D&pool\_size&1&1&\\
\cmidrule{2-6}
&Dropout&rate&0.5&0.5&\\
\cmidrule{2-6}
&Flatten&&&&\\
\midrule
\multirow{3}{*}{\ac{lstm}}&&Units&40&50&40\\
&&Activation&elu&relu&relu\\
\midrule
\multirow{2}{*}{Dense}&&Units&3&CF&CB\\
&&Activation&elu&softmax&softmax\\

\bottomrule
&\\
         \toprule
&&lr&0.0005&0.0001&0.0001\\
\multicolumn{2}{c}{Training parameters}&epochs&100&100&100\\
&&batch\_size&256&100&100\\
&&Optimizer&Adam&Adam&Adam\\
         \bottomrule
         
    \end{tabular}
    \begin{tablenotes}
        \item CF: Number of classes - Floors, CB: Number of classes - Buildings
        \end{tablenotes}
    \end{threeparttable}
\end{table}

%% file: 04-experiments-results.tex
\section{Experiments and Results}
\label{sec:usecases}
\subsection{Experiment setup}

The experiments were executed on a computer with Fedora Linux 32, Intel® Core™ i7-8700T processor and 16 GB of RAM. The methods were implemented in Python 3.9.


A total of fifteen public \ac{wifi} and two \ac{ble} fingerprint datasets were used to test the proposed framework for indoor positioning and data augmentation: UJI~1--2 (UJIIndoorLoc),
LIB~1--2, 
UJIB~1--2
(Universitat Jaume I); TUT~1-7 
(Tampere University); DSI~1-2
(University of Minho); MAN~1-2 
(University of Mannheim) and UTSIndoorLoc (UTS)
(University of Technology Sydney)~\cite{torres2020comprehensive, song2019novel, mendoza2019rss}. All datasets were independently collected, ensuring the experiments are based on heterogeneous datasets and the results can be generalised. Three datasets were used to fine-tune the proposed model (hyperparameters election) and the remaining ones were used to assess the general performance.

Since the data representation may influence the performance of machine learning models, we have selected \textit{powed} data representation (see Eq.~\ref{eq:powed}) to minimize the signal fluctuation existing in the datasets prior training the proposed positioning and data augmentation model. 
\begin{equation}
    \label{eq:powed}
    \begin{aligned}
     Powed & =\begin{cases} 
        0, & \text{if $\ac{rss}_{i}=0$}, \\
        \left(\frac{\ac{rss}_i - min(\Psi)}{-min(\Psi)} \right)^\beta, & \text{otherwise}
    \end{cases}
    \end{aligned}
\end{equation}
where $i$ is the i-$th$ \ac{rss} value, $min(\Psi)$ is the lowest \ac{rss} value in the dataset and $\beta$ is the mathematical constant $e$.

\begin{figure}[!hbt]
    \centering
    \includegraphics[width=0.95\columnwidth,trim={0.25cm 0.2cm 1.25cm 0.8cm},clip]{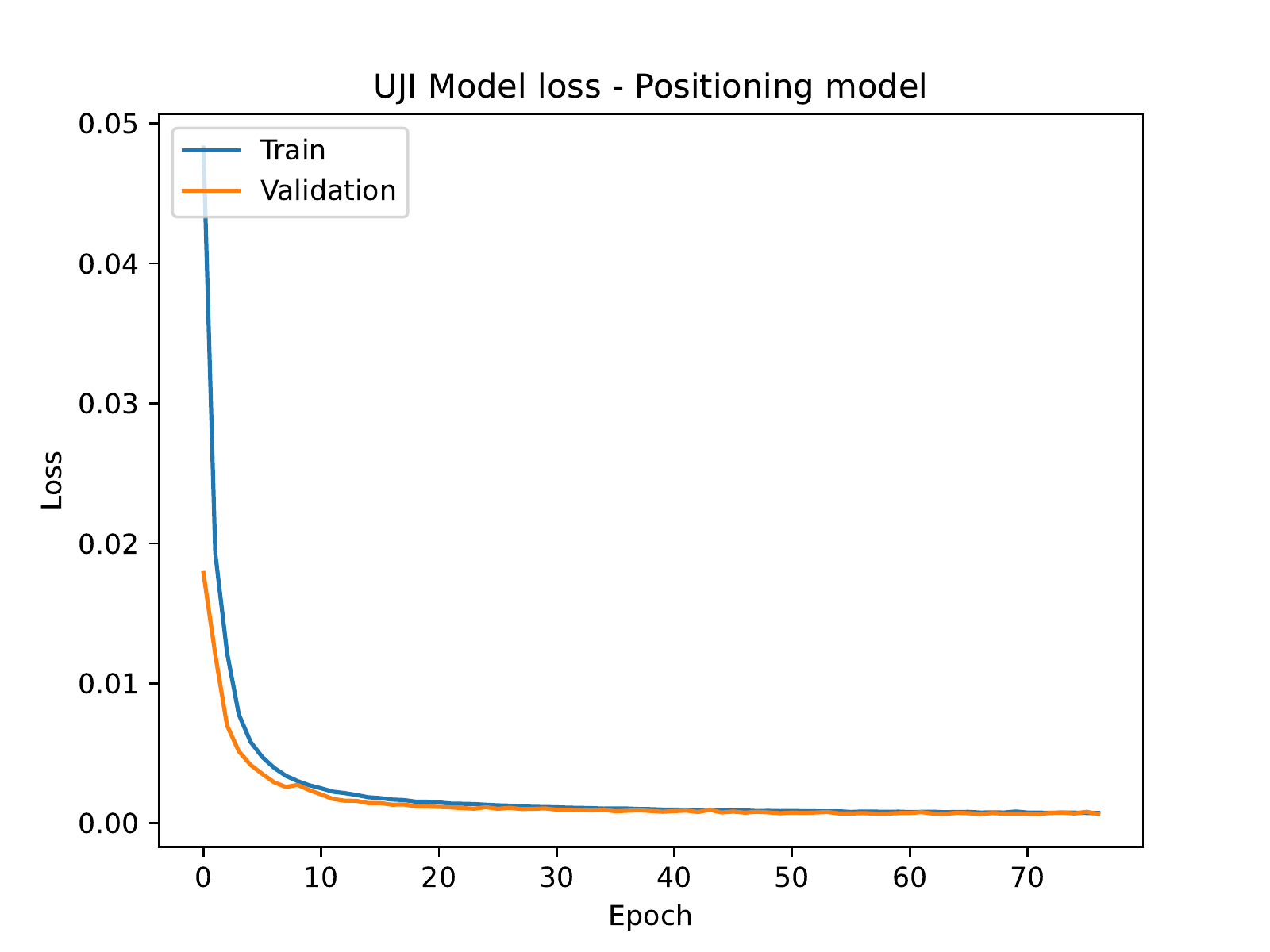}
    \caption{CNN-LSTM model}
    \label{fig:latitudeloss}
\end{figure}

Once the data is normalized, the positioning model (\ac{cnn}-\ac{lstm}) is trained to predict the user or device position. To avoid the overfitting and overtraining issues the \textit{early stopping} method has been used during the training stage. Fig.~\ref{fig:latitudeloss} shows the training loss (blue) vs. the validation loss (orange) of the positioning model for the UJI~1 dataset. 

The first estimator  of the proposed framework (with the original radio map) was tested in terms of 2D and 3D mean positioning error ($\epsilon_{2D}$ and $\epsilon_{3D}$) and floor hit rate ($\gamma$). It is compared to benchmark methods proposed in the literature 
by \textcite{song2019novel} (see Table~\ref{table:comparison3}).

\input{tables/tbl_results}

The results in Table \ref{table:comparison3} show that the proposed \ac{cnn}-\ac{lstm} provides a slightly reduction of the positioning error with respect to CNNLoc~\cite{song2019cnnloc} in UJI~1 and UTS datasets, but the error increased in the TUT~3 dataset. The results of our \ac{cnn}-\ac{lstm} estimator are promising even without data augmentation.

\subsection{Setting the hyperparameters for the GAN network}
\label{subsection:exp_refdatasets}

The proposed \ac{cnn}-\ac{lstm} is working for position estimation using deep learning. However, we consider that its performance can be improved by augmenting the radio map. Thus, we have used three reference datasets found in the literature (UJI~1, TUT~3 and UTS) to select the optimal hyperparameters values to train the proposed cGAN model.

Table \ref{table:results_training} shows the training parameters for the c\ac{gan} architecture and the results obtained with each method and configuration. The number of epochs for each test was set to $14$, and the batch size to $64$ and $128$. The number of iterations is $10$, and the distance between the real samples and the \emph{artificial} samples is between \SIrange{1}{5}{\meter} or \SIrange{1}{10}{\meter}. It means that $200$ samples were generated for each meter, and only those new fingerprints in the defined range were selected.

\input{tables/tbl_gan_training}

\input{tables/tbl_full_comparison_SI}

\subsubsection{Results -- First Method}

M1 was tested only in the UJI~1 dataset due to that this dataset contains samples collected in $3$ buildings and 4--5 floors per building. The first configuration used was the following: batch size equal to $64$, and distance between \SIrange{1}{5}{\meter}. As a result, the positioning error was reduced by $11\%$ in comparison with the error reported by \cite{song2019novel}. When the maximum distance between real and synthetic fingerprints is set to \SI{10}{\meter} the error slightly increased by $6\%$ approximately with regards to the previous configuration.

In the second configuration ($batch size=128$ and distance between \SIrange{1}{5}{\meter}) the results obtained are less accurate than the first configuration increasing the mean positioning error by more than $6\%$. Similarly, the mean positioning error also increased (more than \SI{1}{\meter}) when the maximum distance between real and synthetic fingerprints is \SI{10}{\meter}.

\begin{figure}[!hbt]
    \centering
    \includegraphics[width=0.88\columnwidth,trim={0.15cm 0.15cm 0.13cm 0.15cm},clip]{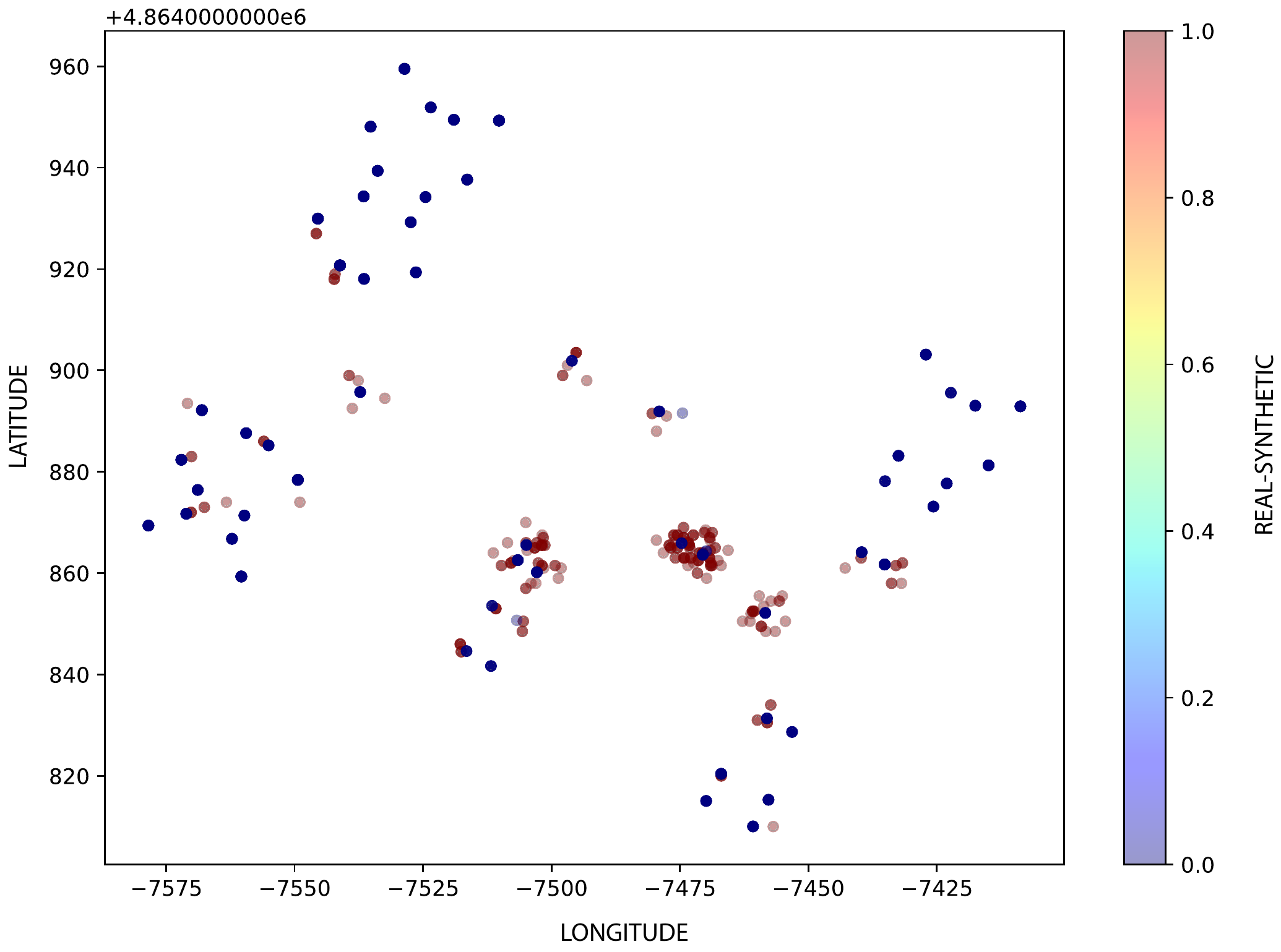}
    \caption{Location of the reference fingerprints for the UJI~1 dataset after applying our generative model (Building 1 and Floor 0).}
    \label{fig:uji-b1f0}
\end{figure}

Fig. \ref{fig:uji-b1f0} illustrates a real example with the position of the reference fingerprints collected empirically (blue dots) and the new fingerprints generated with the  \ac{gan} (red dots) for the UJI~1 datasets. The c\ac{gan} was trained per each building (as defined in the first method) and the distance between the real and new fingerprints is less than \SI{5}{\meter}. 

\subsubsection{Results -- Second Method}

The mean positioning error obtained after training the \ac{gan} network by floor were less than the benchmark. However, it was higher than the first method in the case of the UJI~1 dataset (see Table~\ref{table:results_training}). 

Similarly, the mean positioning error was reduced by $6\%$ in TUT~3 dataset using $14$ epochs, batch size equal to 64 and distance from the real samples is from \SIrange{1}{5}{\meter}. If the distance between samples increased up to \SI{10}{\meter}, the positioning error decreased by more than $11\%$. In the case of UTS, the positioning error was reduced by $5\%$. 

\subsubsection{Results -- Third Method}

The third method provides lower performance than the first and second methods. However, the positioning error is still lower than the benchmark. For instance, the positioning error was reduced by $8\%$ approximately in the UJI~1 dataset.  

In the case of TUT~3 dataset, $3951$ new \emph{artificial} fingerprints were created with the following training configuration: $14$ epochs, batch size equal to $64$ and distance from the real fingerprints between \SIrange{1}{5}{\meter}; thus, the mean 2D position error was reduce by $11\%$ (approx.), from \SI{10.88}{\meter} to \SI{9.67}{\meter} in comparison with the mean 2D positioning error reported by \cite{song2019cnnloc}. When the \ac{gan} is trained with $14$ epochs, the batch size $128$, and the distance less than \SI{10}{\meter}, the positioning error increased by \SI{10}{\centi\meter} despite the fact that the number of artificial fingerprints was more than the previous configuration.
 
Additionally, the c\ac{gan} network was trained using a batch size equal to $128$, $14$ epochs and distance was from \SIrange{1}{5}{\metre}. If we compare the error obtained with the same parameters but the batch size equal to $64$, the error increase a few centimetres (\SI{11}{\centi\metre} approx.) in both 2D and 3D positioning.

\subsection{Generalized results}
In Section \ref{subsection:exp_refdatasets} multiple tests have been done in $3$ reference \ac{wifi} fingerprint datasets in order to get the c\ac{gan} training hyperparameters. In general, the three methods (M1--M3)  have good performance using the batch size equal to $64$ and distance between  \SIrange{1}{5}{\meter}. These parameters were tested in $14$ additional public datasets in order to determine if the suggested configuration can be used in multiple datasets. Unlike the previous experiments, only M2 and M3 were used in the additional datasets, as none of them is multi-building.

To compare the proposed framework, we used the \ac{knn} algorithm as the baseline positioning method (see Table~\ref{table:comparisonknn}). We thus report the normalized values of the 2D and 3D mean positioning error ($\tilde\epsilon_{2D}$ and $\tilde\epsilon_{3D}$) to enhance the relative impact on each dataset~\cite{torressospedra2021ubiquitous}.

\subsubsection{Results - M2} After applying the second method the mean positioning error was reduced by more than $20\%$ in DSI~1, LIB~2 and TUT~1 datasets compared with the \ac{knn}. However, there are some cases where the positioning error increased with regards to the \ac{cnn}-\ac{lstm}, but, it is not higher than $6\%$. Similarly,
the 2D positioning error increased in TUT~2-7 datasets, but the 3D positioning was slightly reduced in TUT~2 dataset. Furthermore, the positioning model improved the floor hit rate by $3\%$ on average (see Table~\ref{table:comparisonknn}).

As mentioned in Section \ref{sec:related_work}, \cite{njima2021indoor} tested their data augmentation model in the UJIIndoorLoc dataset (UJI 1--2) to improve localisation accuracy. In this case, the authors used the data collected in the first building and second floor. Then, the authors selected $1000$ random samples of $1395$ from the training dataset, and the remaining samples were added to the testing dataset. As a result, the authors reduced the positioning error by $15.36\%$ using $1000$ new positions. Following the same procedure but without reducing the number of \acp{ap} in the selected data, our approach reduced the positioning error by more than $19\%$ using $8000$ new synthetic fingerprints.

\subsubsection{Results - M3} This method provides better performance in terms of floor hit rate than the second method. However, the mean 2D and 3D positioning errors were slightly affected in most of the datasets, for instance, the mean 3D positioning error increased by $6\%$ approximately in DSI~1 dataset (see Table~\ref{table:comparisonknn}).

%% file: tables/tbl_results.tex
\begin{table}[!hbtp]
    \tabcolsep 5pt
    \caption{Benchmark and Positioning Model(CNN-LSTM)}
    \label{table:comparison3}
    \centering
    \begin{threeparttable}
    \begin{tabular}{l
    S[table-format=5.0]S[table-format=3.0]S[table-format=2.0]
    S[table-format=2.2]S[table-format=0.0]
    S[table-format=2.2]S[table-format=2.2]
    }
         \toprule
            &\multicolumn{3}{c}{Dim.}
            &\multicolumn{2}{c}{CNNLoc \cite{song2019cnnloc}}
            &\multicolumn{2}{c}{\ac{cnn}-\ac{lstm}}
            \\
        \cmidrule(rl){2-4} 
         \cmidrule(rl){5-6} 
         \cmidrule(rl){7-8} 
            {\multirow{2}{*}{{}Database}} 
            &{\multirow{2}{*}{{$\abs{\tau}$}}}
            &{\multirow{2}{*}{{$\abs{A}$}}}
            &{\multirow{2}{*}{{$\delta_{fp}$}}}
            &{$\epsilon_{2D}$}
            &{$\epsilon_{3D}$}

            &{$\epsilon_{2D}$}
            &{$\epsilon_{3D}$}\\
            &&&&{$[\si{\meter}]$}&{$[\si{\meter}]$}&{$[\si{\meter}]$}&{$[\si{\meter}]$}\\
         \midrule
\textasteriskcentered UJI~1 &20982&520&{\SIrange[range-phrase=--]{1}{77}{}}&11.78&{--}&11.17&11.24\\
\textbullet TUT~3&3951&992&{\SIrange[range-phrase=--]{1}{2}{}}&10.88&{--}&11.14&11.23\\
\textbullet UTS&9496&589&{\SIrange[range-phrase=--]{1}{35}{}}&7.60&{--}&7.34&7.57\\

\bottomrule
    \end{tabular}
    \begin{tablenotes}
        \item \textasteriskcentered Multi-building Multi-floor , \textbullet Single-building Multi-floor, $\abs{\tau}$: number of fingerprints, $\abs{A}$: number of \acp{ap}, $\delta_{fp}$: number of fingerprints per reference point, $\epsilon_{2D}$: mean positioning error 2D, $\epsilon_{3D}$: mean positioning error 3D.
      
    \end{tablenotes}
    \end{threeparttable}
\end{table}

%% file: tables/tbl_gan_training.tex
\begin{table}[!hbtp]
    \tabcolsep 2.85pt
    \caption{Training parameters for the cGAN and  results obtained with each method and configuration for three datasets}
    \label{table:results_training}
    \centering
    \begin{tabular}{l
    S[table-format=2.1]
    S[table-format=3.1]
    c
    S[table-format=2.1]
    c
    S[table-format=5.0]
    S[table-format=2.2]
    S[table-format=2.2]}
         \toprule
         &\multicolumn{5}{c}{Training Parameters}
            &\multicolumn{3}{c}{Positioning}

            \\
         \cmidrule(rl){2-6} 
         \cmidrule(rl){7-9} 
            Database
            & {Epoc.}
            & {BS}
            & {Method}
            & {Iter.}
            & {Dist.}
            &$\eta$
            &$\epsilon_{2D}$
            &$\epsilon_{3D}$\\
         \midrule
\multirow{12}{*}{UJI~1} & 14 & 64 & M1 & 10 & 1-5&155&10.01&10.07\\
& 14 & 64 & M1 & 10 &1-10&1415&10.64&10.70\\
& 14 & 64 & M2 & 10 & 1-5& 213 & 10.55 & 10.60\\
& 14 & 64 & M2 & 10 & 1-10& 1398 & 11.18 & 11.25\\
& 14 & 64 & M3 & 10 & 1-5&288&10.85&10.91\\
& 14 & 64 & M3 & 10 & 1-10&11138&11.00&11.06\\
\cmidrule(rl){2-9}
& 14 & 128 & M1 & 10 &1-5&311&10.67&10.73\\
& 14 & 128 & M1 & 10 &1-10&2088&11.80&11.84\\
& 14 & 128 & M2 & 10 &1-5&311&11.25&11.31\\
& 14 & 128 & M2 & 10 &1-10&2088&11.93&11.99\\
& 14 & 128 & M3 & 10 & 1-5&132&10.97&11.02\\
& 14 & 128 & M3 & 10 & 1-10&1677&11.60&11.66\\
\midrule
\multirow{8}{*}{TUT~3} & 14 & 64 & M2 & 10 & 1-5&848&10.23&10.44\\
& 14 & 64 & M2 & 10 & 1-10&5175&9.62&9.72\\
& 14 & 64 & M3 & 10 & 1-5&3951&9.67&9.76\\
& 14 & 64 & M3 & 10 & 1-10&2912&10.13&10.23\\
\cmidrule(rl){2-9}
& 14 & 128 & M2 & 10 & 1-5&600&11.41&11.51\\
& 14 & 128 & M2 & 10 & 1-10&3498&9.26&9.34\\
& 14 & 128 & M3 & 10 & 1-5&449&11.50&11.61\\
& 14 & 128 & M3 & 10 & 1-10&4847&9.77&9.86\\
\midrule
\multirow{2}{*}{UTS} & 14 & 64 & M2 & 10 & 1-5&266&7.21&7.57\\
& 14 & 64 & M3 & 10 & 1-5&389&7.48&7.68\\
         \bottomrule
    \end{tabular}
\end{table}

%% file: tables/tbl_full_comparison_SI.tex
\begin{table*}[!hbtp]
    \tabcolsep 3.75pt
    
    \caption{Comparison among the 1-NN baseline, CNN-LSTM  and the proposed  models. $\phi$ stands for the number of new fingerprints.}
    \label{table:comparisonknn}
    \centering
    \begin{tabular}{
    l
    S[table-format=5.0]S[table-format=3.0]S[table-format=3.0]
    S[table-format=2.2]S[table-format=2.2]S[table-format=1.0]S[table-format=1.0]S[table-format=3.2]
    S[table-format=1.0]S[table-format=1.0]S[table-format=3.2]
    S[table-format=5.0]S[table-format=1.0]S[table-format=1.0]S[table-format=3.2]
    S[table-format=5.0]S[table-format=1.0]S[table-format=1.0]S[table-format=3.2]
    }
         \toprule
            &\multicolumn{3}{c}{Dim.}
            &\multicolumn{5}{c}{Baseline 1-NN}
            &\multicolumn{3}{c}{\ac{cnn}-\ac{lstm}}
            &\multicolumn{4}{c}{SURIMI M2}
            &\multicolumn{4}{c}{SURIMI M3}
            \\
         \cmidrule(rl){2-4}
         \cmidrule(rl){5-9}
         \cmidrule(rl){10-12} 
         \cmidrule(rl){13-16}
         \cmidrule(rl){17-20}
            \multirow{2}{*}{Database}
            &\multicolumn{1}{c}{\multirow{2}{*}{$\abs{\mathcal{T}}$}}
            &\multicolumn{1}{c}{\multirow{2}{*}{$\abs{\mathcal{A}}$}}
            &\multicolumn{1}{c}{\multirow{2}{*}{$\delta_{fp}$}}
            
            &\multicolumn{1}{c}{{$\epsilon_{2D}$}}
            &\multicolumn{1}{c}{$\epsilon_{3D}$}
            &\multicolumn{1}{c}{$\tilde\epsilon_{2D}$}
            &\multicolumn{1}{c}{$\tilde\epsilon_{3D}$}
            &\multicolumn{1}{c}{{$\gamma$}}

            &\multicolumn{1}{c}{$\tilde\epsilon_{2D}$}
            &\multicolumn{1}{c}{$\tilde\epsilon_{3D}$}
            &\multicolumn{1}{c}{{$\gamma$}}
            
            &\multicolumn{1}{c}{\multirow{2}{*}{$\phi$}}
            &\multicolumn{1}{c}{$\tilde\epsilon_{2D}$}
            &\multicolumn{1}{c}{$\tilde\epsilon_{3D}$}
            &\multicolumn{1}{c}{{$\gamma$}}

            &\multicolumn{1}{c}{\multirow{2}{*}{$\phi$}}
            &\multicolumn{1}{c}{$\tilde\epsilon_{2D}$}
            &\multicolumn{1}{c}{$\tilde\epsilon_{3D}$}
            &\multicolumn{1}{c}{{$\gamma$}}\\
            
            &&&
            
            &\multicolumn{1}{c}{$[\si{\meter}]$}
            &\multicolumn{1}{c}{$[\si{\meter}]$}
            &\multicolumn{1}{c}{$[-]$}
            &\multicolumn{1}{c}{$[-]$}
            &\multicolumn{1}{c}{$[\%]$}
            
            &\multicolumn{1}{c}{$[-]$}
            &\multicolumn{1}{c}{$[-]$}
            &\multicolumn{1}{c}{$[\%]$}
            
            &
            &\multicolumn{1}{c}{$[-]$}
            &\multicolumn{1}{c}{$[-]$}
            &\multicolumn{1}{c}{$[\%]$}

            &
            &\multicolumn{1}{c}{$[-]$}
            &\multicolumn{1}{c}{$[-]$}
            &\multicolumn{1}{c}{$[\%]$}\\
         \midrule
         \multicolumn{20}{c}{\ac{wifi} datasets}\\ 
         \midrule
DSI~1&1369&157&6&4.97&4.97&1&1&100.00&0.88&0.88&100.00&25712&0.74&0.74&100.00&24358&0.79&0.79&100.00\\
DSI~2&576&157&\SIrange{2}{3}{}&4.96&4.96&1&1&100.00&1.29&1.29&100.00&24350&0.98&0.98&100.00&23897&1.02&1.02&100.00\\
LIB~1&576&174&12&3.00&3.01&1&1&99.84&0.96&1.02&99.36&3712&0.94&0.95&98.85&54&0.94&0.98&99.07\\
LIB~2&576&197&12&4.00&4.18&1&1&97.67&0.73&0.74&99.42&969&0.76&0.77&98.33&118&0.80&0.86&99.62\\
MAN~1&14300&28&110&2.84&2.84&1&1&100.00&0.87&0.87&100.00&23578&0.87&0.87&100.00&24247&0.89&0.89&100.00\\
MAN~2&1300&28&10&2.47&2.47&1&1&100.00&0.98&0.98&100.00&11478&0.84&0.84&100.00&6505&0.86&0.86&100.00\\
TUT~1 &1476&309&1&8.59&9.55&1&1&90.00&0.88&0.81&93.27&4541&0.86&0.78&89.18&3526&0.83&0.76&91.63\\
TUT~2&584&354&1&13.00&15.11&1&1&71.02&1.05&0.90&90.34&613&1.03&0.89&89.77&45&1.14&0.98&90.91\\
TUT~4&3951&992&1&6.15&6.40&1&1&95.40&1.02&0.98&96.41&5502&1.03&1.00&96.84&184&1.16&1.13&96.70\\
TUT~5&446&489&1&6.39&6.92&1&1&88.39&1.74&1.62&97.56&596&1.74&1.62&98.57&1232&1.49&1.38&99.29\\
TUT~6&3116&652&1&2.07&2.08&1&1&99.95&2.62&2.62&99.90&6329&2.10&2.10&99.39&186&2.60&2.60&99.96\\
TUT~7&2787&801&1&2.23&2.62&1&1&99.12&2.32&1.98&98.06&809&2.18&1.86&98.00&82&2.17&1.86&98.32\\
\midrule
\multicolumn{20}{c}{BLE datasets} \\
\midrule
UJIB~1&1632&24&\SIrange{30}{36}{}&3.07&3.07&1&1&100.00&0.93&0.93&100.00&1990&0.95&0.95&100.00&1990&0.95&0.95&100.00\\
UJIB~2&816&22&24&4.25&4.25&1&1&100.00&0.93&0.93&100.00&1990&0.93&0.93&100.00&1990&0.90&0.90&100.00\\
\midrule
Avg.&&&&&&1&1&95.53&1.25&1.20&97.84&&1.16&1.10&97.34&&1.20&1.15&97.90\\
         \bottomrule
    \end{tabular}
\end{table*}

%% file: 05-discussion.tex
\section{Discussion}
\label{sec:discussion}

In the earlier research, \ac{gan} was applied in multiple areas to generate new data that can be pass as real. One of the most representative examples is the generation of artificial faces or pictures which are indistinguishable from real images. Considering this, our proposal is devoted to generating new realistic fingerprints to increase the number of samples in the radio maps and reduce the positioning error.

The new \emph{artificial} fingerprints contain special characteristics that allows to enrich the radio map and reduce the necessity of collecting new fingerprints in the physical location. Additionally, it can be considered as a manner to update the radio map. However, the new fingerprints can be located far from the seed fingerprint or in unreachable indoor areas. We have proposed an algorithm to keep only those new fingerprints that are relevant for indoor positioning (see Algorithm~\ref{alg:fp_class_balancing}). 

Additionally, the proposed c\ac{gan} architecture allows us to condition the network by adding a label to the data. It diminishes the possibility of having non-relevant fingerprints in the enriched radio map. We found that the set of hyperparameters for the c\ac{gan} (e.g., number of epochs or the batch size, among others) is the key to generating unique \emph{artificial} fingerprints. 

Given that both the \ac{gan} and the positioning model (\ac{cnn}-\ac{lstm}) use convolutional neural networks, it allows extracting important characteristics of the dataset, and even more when the datasets were collected using crowdsourced techniques or when the fingerprints are not collected systematically. 

It is also important to select an appropriate distance between the real fingerprints and the \emph{artificial} fingerprints. According to the experiments, if the average distance between points in the dataset is more than \SI{2}{\meter}, the distance could be between \SIrange{1}{5}{\meter}, but if the average distance is less than \SI{1}{\meter}, it is better to choose a smaller distance, for instance, \SI{1}{\meter} or less. Due to that c\ac{gan} network learns different characteristics of the fingerprints to generate new synthetic data, this may result in a degradation of the accuracy of the classification model. Thus, in multi-floor or multi-building datasets while more \emph{artificial} fingerprints are added to the original dataset less accurate will be the classification model (see Table~\ref{table:comparisonknn}). This issue might be reduced by increasing the complexity of the classification model or using the model trained with the original data.

We can observe the same behaviour in the data augmentation model proposed by \cite{njima2021indoor}; for instance, the performance of the positioning model is better when using $1000$ new samples than using $2000$. In \cite{njima2021indoor}, we can also observe that the authors slightly modify the hyperparameters and/or the number of neurons of the deep neural network implemented to estimate the position. In such a way, the authors achieved similar results with a different number of new synthetic data. 

Unlike the previous work~\cite{njima2021indoor}, we propose a new framework for positioning estimation and data augmentation, which do not require further modification to provide good performance, as shown in Table~\ref{table:comparisonknn}. i.e., our proposed framework requires less fine-tuning effort.

In the same fashion, the accuracy of the proposed framework may vary from one dataset to another (see Table~\ref{table:comparisonknn}). According to the results obtained, we can notice that the accuracy is also related to the number of fingerprints per reference point, the more fingerprints per reference point in the radio map the more the accuracy is improved and vice-versa. i.e., data augmentation works if the radio map in a particular location is rich enough. 

%% file: 06-conclusions.tex
\section{Conclusions}
\label{sec:conclusions}

In this research work, we provide a new framework for indoor positioning, which consists of a positioning model \ac{cnn}-\ac{lstm} and a c\ac{gan} to generate \emph{artificial} but realistic fingerprints. These two models or architectures are based on deep learning techniques that allow us to extract the most relevant characteristic of the datasets. Thus, the proposed architecture or solution is capable of reducing the positioning error in $70\%$ of the public datasets used in this research work, being the maximum reduction in the positioning error $26\%$ and the minimum $2\%$.

The solution proposed was tested in seventeen public datasets in order to verify if both models are generalizing well and can be used with multiple \ac{wifi} fingerprinting datasets. Furthermore, three methods for training \ac{gan} networks have been tested, the first one to train the \ac{gan} network by building, being the conditional label the floor, in the second method, the \ac{gan} is trained by floor and the last method trains all the dataset without splitting the data. The first and second methods provide better results than the last method. However, all the methods reduced the positioning error with respect to the benchmark.

Future work will analyze new optimization algorithms in order to reduce the instability during the training stage. Also, different variants of \ac{gan} and other methods for data augmentation will be tested to provide a complete study of the advantages and disadvantages of each method in relation to fingerprinting indoor positioning.

%% file: 00-main.bib
@ARTICLE{torres2020comprehensive,
author={Torres-Sospedra, Joaquin and Richter, Philipp and Moreira, Adriano and Mendoza-Silva, German and Lohan, Elena-Simona and Trilles, Sergio and Matey-Sanz, Miguel and Huerta, Joaquin},
journal={IEEE Transactions on Mobile Computing}, 
title={A Comprehensive and Reproducible Comparison of Clustering and Optimization Rules in Wi-Fi Fingerprinting}, 
year={2020},
doi={10.1109/TMC.2020.3017176}}

@INPROCEEDINGS{song2019cnnloc,  author={Song, Xudong and Fan, Xiaochen and He, Xiangjian and Xiang, Chaocan and Ye, Qianwen and Huang, Xiang and Fang, Gengfa and Chen, Liming Luke and Qin, Jing and Wang, Zumin},  booktitle={2019 IEEE SmartWorld, Ubiquitous Intelligence   Computing, Advanced   Trusted Computing, Scalable Computing   Communications, Cloud   Big Data Computing, Internet of People and Smart City Innovation},   title={{CNNLoc: Deep-Learning Based Indoor Localization with WiFi Fingerprinting}},   year={2019},  volume={},  number={},  pages={589-595},  doi={10.1109/SmartWorld-UIC-ATC-SCALCOM-IOP-SCI.2019.00139}}

@INPROCEEDINGS{dai2019autonomous,  author={Dai, Shilong and He, Liang and Zhang, Xuebo},  booktitle={2020 ACM/IEEE 11th International Conference on Cyber-Physical Systems (ICCPS)},   title={Autonomous WiFi Fingerprinting for Indoor Localization},   year={2020},  volume={},  number={},  pages={141-150},  doi={10.1109/ICCPS48487.2020.00021}}

@inproceedings{borras2016indoor,
  title={Indoor positioning with probabilistic WkNN Wi-Fi fingerprinting},
  author={Santiago De Nadal Borras},
  year={2016}
}

@Article{alshami2017adaptive,
AUTHOR = {Alshami, Iyad Husni and Ahmad, Noor Azurati and Sahibuddin, Shamsul and Firdaus, Firdaus},
TITLE = {Adaptive Indoor Positioning Model Based on WLAN-Fingerprinting for Dynamic and Multi-Floor Environments},
JOURNAL = {Sensors},
VOLUME = {17},
YEAR = {2017},
NUMBER = {8},
ARTICLE-NUMBER = {1789},
URL = {https://www.mdpi.com/1424-8220/17/8/1789},
ISSN = {1424-8220},
DOI = {10.3390/s17081789}
}

@ARTICLE{belmonte2020recurrent,  
author={Belmonte-Hernández, Alberto and Hernández-Peñaloza, Gustavo and Martín Gutiérrez, David and Álvarez, Federico},  
journal={IEEE Sensors Journal},   
title={Recurrent Model for Wireless Indoor Tracking and Positioning Recovering Using Generative Networks},   
year={2020},  
volume={20},  
number={6},  
pages={3356-3365},  doi={10.1109/JSEN.2019.2958201}}

@ARTICLE{li2021af,  
author={Li, Qiyue and Qu, Heng and Liu, Zhi and Zhou, Nana and Sun, Wei and Sigg, Stephan and Li, Jie},  journal={IEEE Transactions on Emerging Topics in Computational Intelligence},   title={AF-DCGAN: Amplitude Feature Deep Convolutional GAN for Fingerprint Construction in Indoor Localization Systems},   
year={2021},  
volume={5},  
number={3},  
pages={468-480},  
doi={10.1109/TETCI.2019.2948058}}

@ARTICLE{song2019novel,
author={Song, Xudong and Fan, Xiaochen and Xiang, Chaocan and Ye, Qianwen and Liu, Leyu and Wang, Zumin and He, Xiangjian and Yang, Ning and Fang, Gengfa},
journal={IEEE Access}, 
title={A Novel Convolutional Neural Network Based Indoor Localization Framework With WiFi Fingerprinting}, 
year={2019},
volume={7},
number={},
pages={110698-110709},
doi={10.1109/ACCESS.2019.2933921}}

@ARTICLE{njima2021indoor,  
author={Njima, Wafa and Chafii, Marwa and Chorti, Arsenia and Shubair, Raed M. and Poor, H. Vincent},
journal={IEEE Access},   
title={Indoor Localization Using Data Augmentation via Selective Generative Adversarial Networks},   year={2021},  
volume={9},  
number={},  
pages={98337-98347},  doi={10.1109/ACCESS.2021.3095546}}

@misc{mirza2014conditional,
title={Conditional Generative Adversarial Nets}, author={Mehdi Mirza and Simon Osindero},
year={2014},
eprint={1411.1784},
archivePrefix={arXiv},
primaryClass={cs.LG}
}

@INPROCEEDINGS{lu2018indoor,  
author={lu, Huang and Xingli, Gan and shuang, Li and heng, Zhang and Yaning, Li and Ruihui, Zhu},  
booktitle={2018 Ubiquitous Positioning, Indoor Navigation and Location-Based Services (UPINLBS)},   
title={Indoor Positioning Technology based on Deep Neural Networks},   
year={2018},  
doi={10.1109/UPINLBS.2018.8559721}}

@ARTICLE{yan2021elm,  
author={Yan, Jun and Cao, Yiming and Kang, Bin and Wu, Xiaohuan and Chen, Liang},  journal={IEEE Sensors Journal},   
title={An ELM-Based Semi-Supervised Indoor Localization Technique With Clustering Analysis and Feature Extraction},   year={2021},  
volume={21},  
number={3},  
pages={3635-3644},  
doi={10.1109/JSEN.2020.3028579}}

@INPROCEEDINGS{torres2020new,  
author={Torres-Sospedra, Joaquín and Quezada-Gaibor, Darwin and Mendoza-Silva, Germán M. and Nurmi, Jari and Koucheryavy, Yevgeni and Huerta, Joaquín},  
booktitle={Int. Conf. on Localization and GNSS},   
title={New Cluster Selection and Fine-grained Search for k-Means Clustering and Wi-Fi Fingerprinting},   
year={2020},  
doi={10.1109/ICL-GNSS49876.2020.9115419}}

@ARTICLE{flueratoru2021highaccuracy,
  author={Flueratoru, Laura and Wehrli, Silvan and Magno, Michele and Lohan, Elena Simona and Niculescu, Dragos},
  journal={IEEE Internet of Things Journal}, 
  title={High-Accuracy Ranging and Localization with Ultra-Wideband Communications for Energy-Constrained Devices}, 
  year={2021},
  volume={},
  number={},
  pages={1-1},
  doi={10.1109/JIOT.2021.3125256}}

@ARTICLE{dinh2020smartphone,  
author={Dinh, Thai-Mai Thi and Duong, Ngoc-Son and Sandrasegaran, Kumbesan},  
journal={IEEE Sensors Journal},   
title={Smartphone-Based Indoor Positioning Using BLE iBeacon and Reliable Lightweight Fingerprint Map},   
year={2020},  
volume={20},  
number={17},  
pages={10283-10294},  
doi={10.1109/JSEN.2020.2989411}}

@INPROCEEDINGS{radhakrishnan2015smartphones,
  author={Radhakrishnan, Meera and Misra, Archan and Balan, Rajesh Krishna and Lee, Youngki},
  booktitle={2015 IEEE 12th International Conference on Mobile Ad Hoc and Sensor Systems}, 
  title={Smartphones and BLE Services: Empirical Insights}, 
  year={2015},
  volume={},
  number={},
  pages={226-234},
  doi={10.1109/MASS.2015.92}}

@ARTICLE{li2018vlc,  
author={Li, Yiwei and Ghassemlooy, Zabih and Tang, Xuan and Lin, Bangjiang and Zhang, Yi},  
journal={IEEE Photonics Technology Letters},   
title={A VLC Smartphone Camera Based Indoor Positioning System},   
year={2018},  
volume={30},  
number={13},  
pages={1171-1174},  
doi={10.1109/LPT.2018.2834930}}

@ARTICLE{bai2020dl,  
author={Bai, Siqi and Yan, Mingjiang and Wan, Qun and He, Long and Wang, Xinrui and Li, Junlin},  
journal={IEEE Sensors Journal},   
title={DL-RNN: An Accurate Indoor Localization Method via Double RNNs},   
year={2020},  
volume={20},  
number={1},  
pages={286-295},  
doi={10.1109/JSEN.2019.2936412}}

@ARTICLE{chen2020novel,  
author={Chen, Z. and AlHajri, M. I. and Wu, M. and Ali, N. T. and Shubair, R. M.},  
journal={IEEE Sensors Letters},   
title={A Novel Real-Time Deep Learning Approach for Indoor Localization Based on RF Environment Identification},   year={2020},  
volume={4},  
number={6},  
pages={1-4},  
doi={10.1109/LSENS.2020.2991145}}

@misc{xu2015empirical,
title={Empirical Evaluation of Rectified Activations in Convolutional Network}, 
author={Bing Xu and Naiyan Wang and Tianqi Chen and Mu Li},
year={2015},
eprint={1505.00853},
archivePrefix={arXiv},
primaryClass={cs.LG}
}

@article{torres2015comprehensive,
author = {Torres-Sospedra, Joaquín and Montoliu, Raúl and Trilles Oliver, Sergi and Belmonte Fernández, Oscar and Huerta, Joaquín},
year = {2015},
month = {12},
pages = {9263-9278},
title = {Comprehensive Analysis of Distance and Similarity Measures for Wi-Fi Fingerprinting Indoor Positioning Systems},
volume = {42},
journal = {Expert Systems with Applications},
doi = {10.1016/j.eswa.2015.08.013}
}

@INPROCEEDINGS{mao2018scalability,
  author={Mao, Yingling and Liu, Ke and Li, Hao and Tian, Xiaohua and Wang, Xinbing},
  booktitle={2018 15th Annual IEEE International Conference on Sensing, Communication, and Networking}, 
  title={Scalability of Wireless Fingerprinting Based Indoor Localization Systems}, 
  year={2018},
  doi={10.1109/SAHCN.2018.8397112}}

@Article{mendoza2019rss,
AUTHOR = {Mendoza-Silva, Germán Martín and Matey-Sanz, Miguel and Torres-Sospedra, Joaquín and Huerta, Joaquín},
TITLE = {BLE RSS Measurements Dataset for Research on Accurate Indoor Positioning},
JOURNAL = {Data},
VOLUME = {4},
YEAR = {2019},
NUMBER = {1},
ARTICLE-NUMBER = {12},
URL = {https://www.mdpi.com/2306-5729/4/1/12},
ISSN = {2306-5729},
DOI = {10.3390/data4010012}
}

@misc{quezada2021surimi,
author = {Quezada-Gaibor, Darwin and Torres-Sospedra, Joaquín and Nurmi, Jari and Koucheryavy, Yevgeni and Huerta, Joaquín},
  title = {{SURIMI: Supervised Radio Map augmentation withDeep Learning and a Generative AdversarialNetwork for Fingerprint-based Indoor Positioning -- Source code}},
  year = {2021},
  publisher = {GitHub},
  journal = {GitHub repository},
  howpublished = {\url{https://github.com/darwinquezada/SURIMI}},
  commit = {4f57d6a0e4c030202a07a60bc1bb1ed1544bf679}
}

@INPROCEEDINGS{asmar2004smartslam,
  author={Asmar, D.C. and Zelek, J.S. and Abdallah, S.M.},
  booktitle={2004 IEEE Int. Conf.on Systems, Man and Cybernetics}, 
  title={SmartSLAM: localization and mapping across multi-environments}, 
  year={2004},
  volume={6},
  number={},
  pages={5240-5245 vol.6},
  doi={10.1109/ICSMC.2004.1401026}}

@inproceedings{miura2006adequate,
author = {Miura, Hirokazu and Sakamoto, Junichi and Matsuda, Noriyuki and Taki, Hirokazu and Abe, Noriyuki and Hori, Satoshi},
year = {2006},
month = {10},
pages = {628-636},
booktitle={Knowledge-Based Intelligent Information and Engineering Systems},
title = {Adequate RSSI Determination Method by Making Use of SVM for Indoor Localization},
volume = {4252},
isbn = {978-3-540-46537-9},
doi = {10.1007/11893004_81}
}

@INPROCEEDINGS{torressospedra2021ubiquitous,
  author={Torres-Sospedra, Joaquín and Silva, Ivo and Klus, Lucie and Quezada-Gaibor, Darwin and Crivello, Antonino and Barsocchi, Paolo and Pendão, Cristiano and Lohan, Elena Simona and Nurmi, Jari and Moreira, Adriano},
  booktitle={2021 International Conference on Indoor Positioning and Indoor Navigation (IPIN)}, 
  title={Towards Ubiquitous Indoor Positioning: Comparing Systems across Heterogeneous Datasets}, 
  year={2021},
  volume={},
  number={},
  pages={1-8},
  doi={10.1109/IPIN51156.2021.9662560}}

@INPROCEEDINGS{radar,
  author={Bahl, P. and Padmanabhan, V.N.},
  booktitle={Proceedings IEEE INFOCOM 2000}, 
  title={{RADAR: an in-building RF-based user location and tracking system}}, 
  year={2000},
  volume={2},
  number={},
  pages={775-784 vol.2},
  doi={10.1109/INFCOM.2000.832252}}
